\documentclass[smallcondensed]{svjour3}       
\usepackage{graphicx}
\usepackage{amsmath}
\usepackage{amssymb}
\usepackage{booktabs} 
\usepackage{subfig}
\usepackage[colorlinks=true, linkcolor=blue, citecolor=blue, urlcolor=black]{hyperref}
\providecommand{\e}[1]{\ensuremath{\times 10^{#1}}} 
\AtBeginDocument{%
	\setlength{\oddsidemargin}{\dimexpr(\paperwidth-\textwidth)/2-1in}%
	\setlength{\evensidemargin}{\oddsidemargin}%
	\setlength{\topmargin}{%
		\dimexpr(\paperheight-\textheight)/2-\headheight-\headsep-1in}%
}

\newcommand{\ket}[1]{\mathinner{|{#1}\rangle}}
\newcommand{\bra}[1]{\mathinner{\langle {#1}|}}

\def\makeheadbox{\relax}
\begin{document}

\title{Entangled photon added coherent states
}


\author{Francisco A. Dom\'inguez-Serna  \and Francisco J. Mendieta-Jimenez \and
       Fernando Rojas 
}


\institute{F. A. Domínguez Serna \at
              Posgrado en F\'isica de Materiales, Centro de Investigaci\'on Cient\'ifica y de Educaci\'on Superior de Ensenada, Ensenada 22890, Baja California, M\'exico.
              \email{francisco.ds@gmail.com}           
			\and F. J. Mendieta-Jimenez \at
				Agencia Espacial Mexicana, Xola y Universidad, D.F., 03020, M\'exico\\
				\email{mendieta.javier@aem.gob.mx}
           \and
           F. Rojas \at
              	Centro de Nanociencias y Nanotecnolog\'ia – Departamento de F\'isica Te\'orica,  Universidad Nacional Aut\'onoma de M\'exico, Ensenada, Baja California, 22860 M\'exico\\
              	\email{frojas@cnyn.unam.mx}
}
\date{Published in 2016. "The final publication is available at Springer via http://dx.doi.org/10.1007/s11128-016-1325-9".	}

\maketitle

\begin{abstract}
We study the degree of entanglement of arbitrary superpositions of $m,n$ photon-added coherent states
(PACS) $\ket{\psi} \propto u \ket{{\alpha},m}\ket{{\beta},n }+ v \ket{{\beta},n}\ket{{\alpha},m}$ using the concurrence,
and obtain the general conditions for maximal entanglement. We show that photon addition process can be identified as an entanglement
enhancer operation for superpositions of coherent states (SCS). Specifically for the known bipartite positive SCS:
$\ket{\psi} \propto \ket{\alpha}_a\ket{-\alpha}_b + \ket{-\alpha}_a\ket{\alpha}_b $ whose entanglement tends to zero for $\alpha \to 0$, can  be
 maximal if al least one photon is  added in a subsystem. A full family of maximally entangled PACS is also presented. We also analyzed the decoherence effects in the entangled PACS induced by a simple depolarizing channel.  We find that robustness against depolarization is increased by adding photons
to the coherent states of the superposition.
We obtain the  dependence of the critical depolarization  $p_{\text{crit}}$ for null entanglement
as a   function of $m,n, \alpha$ and  $\beta $.

\end{abstract}

\section{Introduction}

\label{intro}
Quantum information processing (QIP) uses the inherent properties of quantum systems like the entanglement  which has been widely considered as a useful
 resource to perform quantum operations, universal quantum computing (QC) and quantum communications \cite{Jozsa2003,Hughes2007,Bennett2000}. Single photon sources
 could generate highly pure and highly entangled single photon states and have been extensively studied for quantum information communication
 since Knill et al.; contribution \cite{Knill01a}. In addition, continuous variable entangled states, play also an important role in performing quantum communication
 protocols and quantum computation tasks \cite{Wang2015,Madsen2012,Neergaard-Nielsen2010,Braunstein2005} (and references therein). The study and generation of
 entangled coherent states is of much interest, since coherent states are  macroscopic and  classical states that can be easily obtained from the known laser sources.
 Therefore, entangling macroscopic states, opens wide possibilities of implementing QIP tasks in the macroscopic world using quantum properties of light.

Photon Added Coherent States (PACS) share properties with purely classical coherent states (CS) $\ket{\alpha}$ and pure quantum Fock states $\ket{n}$ and
 have been located in between the classical-quantum regime \cite{Zavatta2004}.PACS are interesting because they present sub-Poissonian statistics in photon number,
 which is related to a more deterministic behavior in the photon counting for a given time window compared with purely classical states whose statistics are Poissonian
 or super-Poissonian \cite{Agarwal1991}. PACS are also characterized by a negativity in the Wigner function, which is a signature of non-classical states.
 This intrinsically makes them more suitable for QIP applications \cite{Kenfack2004}. Nevertheless, photon addition together with photon subtraction are part of the non-Gaussian quantum operations that are useful to generate Gaussian entangled states (GES) or to increase entanglement of an existing GES \cite{Bartley2015,Eisert2002}.

Previous studies of PACS,  in the direction of entangled channels and their  application to quantum communications have been done. For example, in \cite{Pinheiro2012},
 quantum communications protocols are explored by adding equal number of photons to a bipartite superposition of CS with opposite phases in the form $\ket{\psi} \propto \hat{a}^{\dagger m}\otimes \hat{b}^{\dagger m} (\ket{\alpha}_a\ket{\alpha}_b+\ket{-\alpha}_a\ket{-\alpha}_b)$ where $\hat{a}^{\dagger} (\hat{b}^{\dagger})$ is the photon creation operator for mode $\hat{a} (\hat{b})$.
 They start by constructing a qubit in the form $\ket{0}\propto \hat{a}^{\dagger m}\ket{\alpha}$ and $\ket{1}\propto \hat{a}^{\dagger m}\ket{-\alpha}$, and use
 these states to formulate quantum teleportation and quantum key distribution. In \cite{Nogueira2013} the entanglement properties of one same weighted pure bipartite CS
 superposition with different number of photon addition is studied, where the CS have opposite phases and small values of $m$ and $n$ with the structure $\ket{\psi} \propto \hat{a}^{\dagger m}\otimes \hat{b}^{\dagger n}\ket{\alpha}_a\ket{-\beta}_b+ \hat{a}^{\dagger n}\otimes \hat{b}^{\dagger m}\ket{-\beta}_a\ket{\alpha}_b$. However a full characterization of
 the entanglement properties in both pure and mixed systems is still needed to evaluate the effect of an arbitrary superposition with different coherent states and photon added numbers.
We study the effect of decoherence by a general depolarizing channel
and the relation of a critical depolarization probability that produces null entanglement in function of the parameters $m,n, \alpha$ and  $\beta $.

In this paper we characterize the entanglement properties of arbitrary superpositions of states of the form
$\ket{\psi} \propto u \ket{{\alpha},m}\ket{{\beta},n }+ v \ket{{\beta},n}\ket{{\alpha},m}$ where $\ket{{\alpha},m}$ stands for an $m$ photon added coherent state up to
a normalization constant in the form $\ket{\alpha,m}\equiv\hat{a} ^{\dagger m (n)}\ket{\alpha}$ and similarly for $\ket{{\beta},n}$.
We first study the general conditions, in the pure state, to obtain maximally entangled photon added states and compare them with the entanglement of the known
entangled coherent state (ECS) of the form $\ket{\psi}\propto \ket{\alpha}\ket{-\alpha} - \ket{-\alpha}\ket{\alpha}$ where the photon addition operation is
 used to enhance entanglement properties. In the second part, we consider a mixed state where the PACS superpositions undergo the simplest model of a depolarization channel. We study the critical values
  for null entanglement as a  function of probability weights in the statistical mixture $p$,
 photon added numbers($m,n$) and coherent states $\ket{\alpha}, \ket{\beta}$  of the PACS.

\subsection{Photon Added Coherent States: PACS}
 PACS were first introduced by Agarwal and Tara \cite{Agarwal1991} in the form,
\begin{equation}
 \ket{\alpha,m}=\frac{\hat{a} ^{\dagger m} \ket{\alpha}}{[m! L_m (-|\alpha|^2)]^{1/2}   },
 \label{eq:AgarwalAdded}
\end{equation}
where $\ket{\alpha}$ is a coherent state (CS) and
$L_m (x)$ is the Laguerre polynomial of order m, defined by
\begin{equation}
L_m(x)=\sum\limits_{n=0}^{m}\frac{(-1)^n x^n m!}{(n!)^2 (m-n)!}
\end{equation}

 Coherent States are eigenstates of the annihilation operator $\hat{a}\ket{\alpha}=\alpha\ket{\alpha}$ and $\hat{a}^{\dagger m }$ is the creation
 operator applied $m$ times. For the determination of the  entanglement we require the value of overlap between two non-normalized states with $m(n)$
 added photons to the coherent state $\ket{\alpha} (\ket{\beta})$,  $\hat{a}^{\dagger m}\ket{\alpha}$ and $\hat{a}^{\dagger n}\ket{\beta}$,
 namely $\bra{\alpha}\hat{a}^{m}\hat{a}^{\dagger n}\ket{\beta}$

By expanding in terms of photon numbers $n$ and $m$ for each mode, we have
	\begin{equation}
	\begin{aligned}
	\bra{\alpha}\hat{a}^{m}\hat{a}^{\dagger n}\ket{\beta} &= \sum\limits_{i,j=0}^{\infty} e^{-|\alpha|^2/2} e^{-|\beta|^2/2}\frac{\alpha^{*i}\beta ^j}{ \sqrt{i!}\sqrt{j!} } \bra{i} \hat{a}^m  \hat{a}^{\dagger n} \ket{j}\\
	&=e^{-(|\alpha|^2+|\beta|^2)/2} \sum\limits_{i,j=0}^{\infty} \bra{i+m} \frac{\sqrt{(m+1)!}}{i!} \alpha^{* i} \beta^j \frac{\sqrt{(n+j)!}}{j!}\ket{n+j},
	\end{aligned}
	\end{equation}
Using the orthogonality  of the Fock space $\delta_{i+m,n+j} \Rightarrow j=i+m-n$, for the inner bra-ket, we can write it as
	\begin{equation}
	\begin{aligned}
	\bra{\alpha}\hat{a}^{m}\hat{a}^{\dagger n}\ket{\beta}=e^{-(|\alpha|^2 + |\beta|^2)/2} \sum_{i} \frac{(\alpha^*\beta)^i \beta^{m-n} (m+i)!}{i!(i+m-n)!}
	\end{aligned}
	\end{equation}
	After summation, the final expression is given by
	\begin{equation}
	\begin{aligned}
	\bra{\alpha}\hat{a}^{m}\hat{a}^{\dagger n}\ket{\beta} &= e^{ -(|\alpha|^2 + |\beta|^2)/2  } \beta^{m-n} \Gamma(1+m) {}_1\tilde{F}_1(1+m; 1+m-n;\alpha^*\beta).
	\end{aligned}
	\end{equation}
where ${}_1\tilde{F}_1(a;b;c)$ is the regularized confluent hypergeometric function of the first kind. In terms of the
confluent hypergeometric function ${}_1\tilde{F}_1(a;b;c)={}_1F_1(a;b;c)/\Gamma(b)$.

\section{Entanglement of PACS superpositions }
We are interested in  the properties of the superpositions of PACS as  bipartite states of the form
\begin{equation}
	\begin{aligned}
	\ket{\psi^{AB}}&=N( u \hat{a}^{\dagger m}\ket{\alpha}_A \hat{b}^{\dagger n} \ket{\beta}_B + v \hat{a}^{\dagger n}\ket{\beta}_A\hat{b}^{\dagger  m}\ket{\alpha}_B    )\\
	& = N (u (\hat{a}^{\dagger m}\otimes \hat{b}^{\dagger n} ) \ket{\alpha}_A\ket{\beta}_B + v (\hat{a}^{\dagger n}\otimes \hat{b}^{\dagger m})\ket{\beta}_A\ket{\alpha}_B   ),
	\end{aligned}
	\label{eq:estadoAB}
\end{equation}
where $\hat{a}^{\dagger m} (\hat{b}^{\dagger n})$ creates $m$ ($n$) photons on the mode $\hat{a}$ of subsystem $A$ ($\hat{b}$ on subsystem $B$), $u$ and $v$ are
complex numbers obeying $|u|^2+|v|^2=1$ and $N$ is a normalization factor due to the non-orthogonal properties and is  given in terms of the parameters $(u,v,\alpha,\beta,m,n)$ by
\begin{equation}
	\begin{aligned}
		N(u,v)&=\Big[ L_m(-|\alpha|^2)L_n(-|\beta|^2)m!n!+2 \Re (u^*v) e^{-(|\alpha|^2+|\beta|^2)}|\beta|^{2(m-n)}  \\
		&\quad \times (\Gamma(1+m))^2|{}_1\tilde{F}_1 (1+m;1+m-n;\alpha^*\beta  )|^2   \Big]^{-1/2}.
	\end{aligned}
	\label{eq:normEdoAB}
\end{equation}

We proceed to build an orthogonal two qubit computational basis by means of the Gram-Schmidt orthogonalization process applied to (\ref{eq:estadoAB}), with  the qubit in each subspace  defined as
\begin{equation}
	\begin{aligned}
		&\ket{0}\equiv  N_1 \hat{a}^{\dagger m} \ket{\alpha},\\
		&\ket{1}\equiv  N_2 ( \hat{a}^{\dagger n} \ket{\beta}- z N_1 \hat{a}^{\dagger m}\ket{\alpha}   ),
	\end{aligned}
	\label{eq:GramBasis}
\end{equation}
where $z=N_1^* \bra{\alpha} \hat{a}^m\hat{a}^{\dagger n}\ket{\beta}$ takes into account the overlap between the non-orthogonal states, with the normalization constants $N_1$ and $N_2$ given by
\begin{equation}
	\begin{aligned}
		 N_1&=[L_m(-|\alpha|^2)m!]^{-1/2},\\
		 N_2&=\big[L_n(-|\beta|^2)n! +|z N_1|^2 L_m(-|\alpha|^2)m!\\
		&\quad -2 e^{-(|\alpha|^2+|\beta|^2)/2} \Gamma(1+m) \Re ( z N_1 \beta^{* (m-n)}\\
		&\quad \times {}_1\tilde{F}_1 (1+m;1+m-n;\beta^*\alpha)    )      \big]^{-1/2}.
	\end{aligned}
	\label{eq:N1_N2}
\end{equation}

In  writing Eq. (\ref{eq:estadoAB}) no initial phase to the coherent states $\alpha$ nor $\beta$ is assumed, as we are interested in finding the conditions of entanglement
for arbitrary superpositions of PACS.

The state in Eq. (\ref{eq:estadoAB}) can be written in terms of the two qubit orthogonal computational basis as
\begin{equation}
	\ket{\psi^{AB}}=\frac{N(u,v)}{N_1 N_2} (u\ket{01}+v\ket{10} + z N_2(u+v)\ket{00} ).
	\label{eq:EstadoAB_BaseON}
\end{equation}

To calculate the degree of entanglement we use the Concurrence \cite{Wootters1998} which can be determined for  pure states of Eq. (\ref{eq:EstadoAB_BaseON}) as
 $C=|\bra{\psi^{AB}}\sigma_y \otimes \sigma_y\ket{\psi^{AB *}}| $, where the asterisk denotes complex conjugation in the computational basis in the way $\ket{\psi^{AB *}}=\sum\limits_{\{i,j\}=0}^{1} \ket{i,j}\bra{i,j}  \psi^{AB}\rangle^*$, and using the time inversion operation with the y-Pauli matrices. Therefore for the entangled PACS the concurrence is
\begin{equation}
	C (\psi^{AB}) = 2\left(\frac{N(u,v)}{N_1 N_2}\right)^2 |uv|^2.
	\label{eq:Concurrence_Estado_AB}
\end{equation}

\begin{figure}
\includegraphics[width=0.98\textwidth]{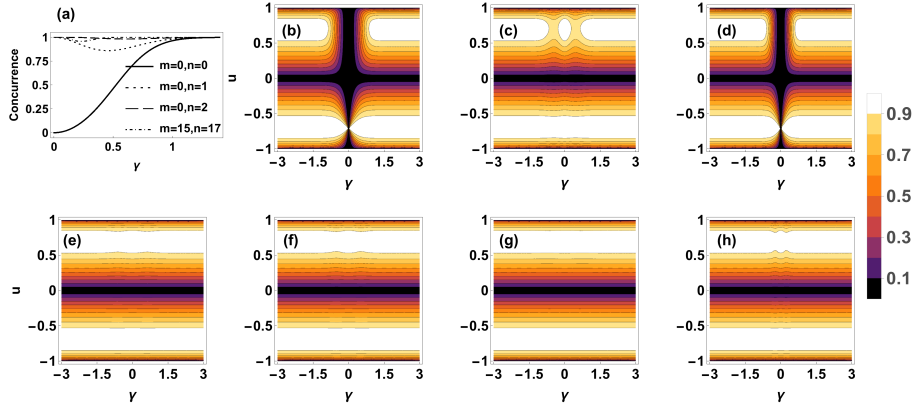}
\caption{Concurrence of Eq. (\ref{eq:edocgamma}) with $\alpha=\beta=0$ for different values of $\gamma$. The conditions considered are for (a) $\ket{\psi} = N( \hat{a}^{\dagger m}\otimes \hat{b}^{\dagger n} \ket{\gamma}\ket{-\gamma} + \hat{a}^{\dagger n}\otimes \hat{b}^{\dagger m} \ket{-\gamma}\ket{\gamma})$ for different $m,n$. For (b)-(h) the state under consideration is $\ket{\psi} = N( u\hat{a}^{\dagger m}\otimes \hat{b}^{\dagger n} \ket{\gamma}\ket{-\gamma} + v\hat{a}^{\dagger n}\otimes \hat{b}^{\dagger m} \ket{-\gamma}\ket{\gamma})$ with (b)  $m=n=0$, (c) $m=0,n=1 $, (d) $m=1,n=1 $, (e) $m=2,n=0$, (f) $m=1,n=3 $, (g) $m=0,n=3 $, (h) $m=15,n=17$.}
\label{fig:comparison}
\end{figure}

A more general superposition can be created rewriting Eq. (\ref{eq:estadoAB}) in the form
\begin{equation}
	\ket{\psi^{AB}} = N (u (\hat{a}^{\dagger m}\otimes \hat{b}^{\dagger n} ) \ket{\alpha + \gamma}_A\ket{\beta-\gamma}_B + v (\hat{a}^{\dagger n}\otimes \hat{b}^{\dagger m})\ket{\beta-\gamma}_A\ket{\alpha+\gamma}_B,
	\label{eq:edocgamma}
\end{equation}
which allows a direct comparison with the known  entangled coherent state (ECS) \cite{Sanders2012,Sanders1992}. We note that when $\alpha=\beta=0$, $u=v=1/\sqrt{2}$, $\gamma \ne 0$ and considering no added photons $n=m=0$, Eq. (\ref{eq:edocgamma}) becomes the known state
\begin{equation}
 \ket{\psi}= N(\ket{\gamma}\ket{-\gamma}\pm \ket{-\gamma}\ket{\gamma} ).
 \label{eq:ECS}
\end{equation}
The entangled properties of this state will be partially used as reference to discuss our results for PACS. It is important to note that,
for the negative superposition in \eqref{eq:ECS} entanglement is maximum even for $|\gamma|$ small, as seen in \eqref{eq:EstadoAB_BaseON}. Concurrence for
the state in Eq. (\ref{eq:edocgamma}) with $\alpha=\beta=0$ for different values of $\gamma$ is shown in Fig. \ref{fig:comparison}.
The states under consideration are as follows: in Fig. \ref{fig:comparison}(a) the state is $\ket{\psi} = N( \hat{a}^{\dagger m}\otimes \hat{b}^{\dagger n} \ket{\gamma}\ket{-\gamma} + \hat{a}^{\dagger n}\otimes \hat{b}^{\dagger m} \ket{-\gamma}\ket{\gamma})$,
where the result for ECS  Eq. (\ref{eq:ECS}) with $m,n=0$ is represented by the solid black line. One can see that the concurrence $C(\psi^{AB})$ increases asymptotically from 0 to 1 as $\gamma$ increases.
On the other hand, if at least one photon is added in one of the modes ($m$ or $n$ larger than zero), concurrence starts from 1. This is because even for null $\gamma$ there is
an entangled Fock photon state. This makes  the entangled PACS more  robust even for a small  photon number in the positive coherent channel. From Fig. \ref{fig:comparison}(b) to \ref{fig:comparison}(h)
the concurrence for the state $\ket{\psi} = N( u\hat{a}^{\dagger m}\otimes \hat{b}^{\dagger n} \ket{\gamma}\ket{-\gamma} + v\hat{a}^{\dagger n}\otimes \hat{b}^{\dagger m} \ket{-\gamma}\ket{\gamma})$ is plotted  as a
 function of $\gamma$ and $u$ for fixed values of added photon number $n,m$. Specifically, in Fig. \ref{fig:comparison}(b) we present $C$  for zero photon added in both channels $m=n=0$,
  in Fig. \ref{fig:comparison}(c) one photon is added with $m=0,n=1 $, the case of two photons added  $m=1,n=1 $ is shown in  Fig. \ref{fig:comparison}(d), two photon added $m=2,n=0$ is
  presented in Fig. \ref{fig:comparison}(e), in Fig. \ref{fig:comparison}(f) the case $m=1,n=3 $, in  Fig. \ref{fig:comparison}(g) $m=0,n=3 $
  and large added photons in Fig. \ref{fig:comparison}(h) $m=15,n=17$. From an analysis of entanglement properties of these results we can conclude that a more robust form of
  entanglement generation can be obtained when $m \ne n$, since  the photon added process decrease the overlap between  the individual states of the bipartite system, i.e.,
   they become more distinguishable and coherent states with a small number of photons (small $\gamma$) still maintain maximum entanglement around the symmetric superposition $u=v=1/\sqrt{2}$.
This non equal photon added  coherent entangled state can be used as a robust and alternative resource channel for the teleportation protocol presented in \cite{Pinheiro2012}

\begin{figure}
	\centering
\includegraphics[width=0.95\textwidth]{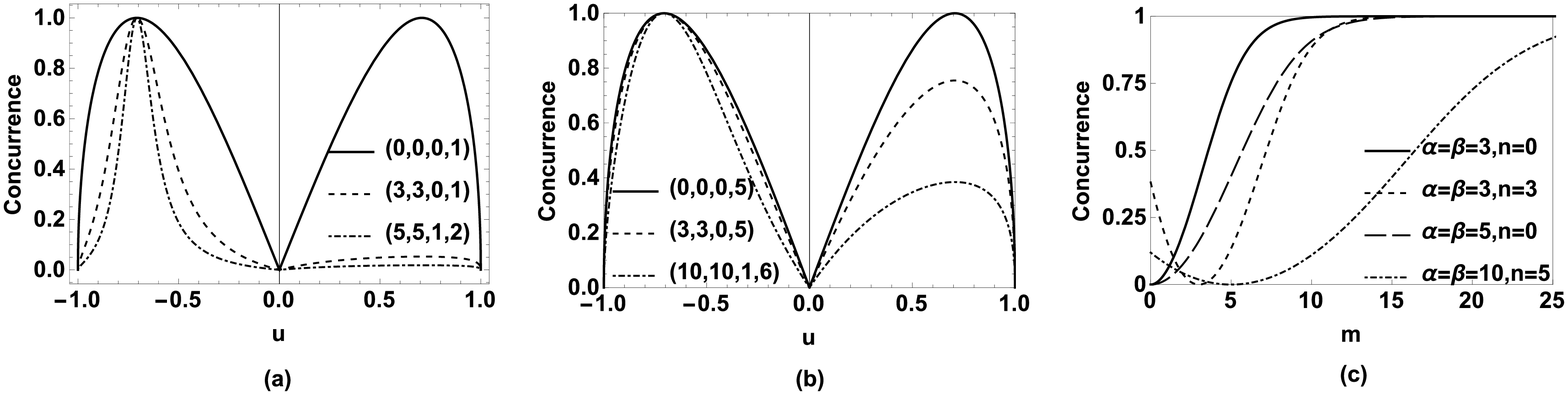}	
	\caption{Concurrence for the state of Eq. (\ref{eq:edocgamma}) for (a) $\alpha=\beta$ fixed, $|m-n|=1$ vs $u$, (b) $\alpha=\beta$ fixed, $|m-n|=5$ vs $u$, (c) $\alpha=\beta$ and $n$ fixed, $u=v=1/\sqrt{2}$ for different numbers of $m$ added photons. Each PACS superposition is characterized with the parameters $(\alpha, \beta, m, n)$.}
  	
\label{fig:ConcG1}
\end{figure}

We  have described the ECS  entangled properties , obtained from the condition  $\alpha=\beta =0$, we now proceed to analyze the  general case with
 $\alpha \ne \beta \ne 0$ and set $\gamma=0$. This is equivalent to absorb $\gamma$ included in the coherent state amplitudes ($\alpha, \beta$).
 The values of concurrence, $C$ are shown in Fig. \ref{fig:ConcG1} for different parameters of the state in Eq. (\ref{eq:EstadoAB_BaseON}).
 In Fig. \ref{fig:ConcG1}(a), we plot concurrence $C$ versus  $u$, where the difference of added photons in each mode  is kept fixed to $|m-n|=1$, while in Fig. \ref{fig:ConcG1}(b) $|m-n|=5$.
 For Fig. \ref{fig:ConcG1}(a) and \ref{fig:ConcG1}(b), the concurrence is obtained for different values of $\alpha=\beta$ vs. $u$ for $u \in \Re$ in the interval $[-1,1]$.
 Two maxima to $C$ are obtained for $u=\pm 1/\sqrt{2}$, where a bell type state is generated. The maximum for the negative superposition $u=-1/\sqrt{2}$ implies $u=-v$ which
is known to be maximally entangled from Eq. (\ref{eq:EstadoAB_BaseON}) where the coefficient of $\ket{00}$ vanishes; on the other hand, the maximum for
the positive superposition $u=v=1/\sqrt{2}$ is $\{\alpha, \beta, |m-n| \}$ dependent. It seems that $C(\psi^{AB}) $ increases along with $|m-n|$ and $(\alpha \beta)^{-1}$. 
More specifically, when $u=v=1/\sqrt{2}$ the concurrence can be increased, by increasing the difference $|m-n|$ for a given pair $\{\alpha,\beta\}$. 
This effect is clearly shown  in Fig. \ref{fig:ConcG1}(c), where the concurrence $C$ for a state with $\alpha=\beta$ is plotted for different values of $|m-n|$ as $n$ 
is kept fixed. Clearly this shows that for a given state one can increase its concurrence by the operation of adding photons. This is the mechanism of non-Gaussian 
operations over coherent states, analog to \cite{Navarrete-Benlloch2012}. This  effect is interesting since without the need of having coherent states with a $\pi$
 phase difference, maximal entangled states as a result of entanglement distillation can be achieved with the addition of photons. Although the states obtained are $\alpha$
  dependent, $C(\psi^{AB})\rightarrow 1$ is obtainable.

\begin{figure}
	\centering
\includegraphics[width=0.95\textwidth]{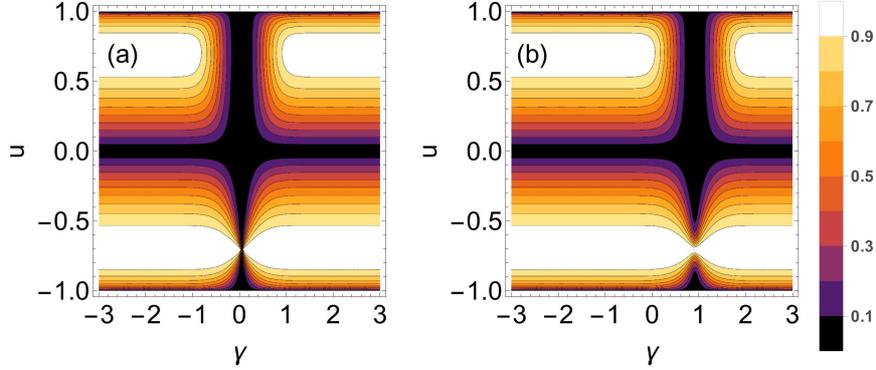}

\caption{Concurrence for the state in Eq. (\ref{eq:edocgamma}) as  a function of $\gamma$ and $u$ (a) $\alpha=\beta=10,m=1,n=2$ , (b) $\alpha=\beta=10,m=0,n=20$ }
	\label{fig:ConcG2}
\end{figure}

In the general case  of state of Eq. (\ref{eq:edocgamma}), all parameters  are non zero.  In Fig. \ref{fig:ConcG2}, concurrence is calculated for a state of the 
form $\ket{\psi} = N ( u \hat{a}^{\dagger m}\otimes \hat{b}^{\dagger n} \ket{\alpha +\gamma}\ket{\alpha - \gamma} + v \hat{a}^{\dagger n} \otimes \hat {b}^{\dagger m} \ket{\alpha -\gamma}\ket{\alpha+\gamma} )$ 
with fixed $\alpha=10$ . Concurrence $C$ is shown in Fig. \ref{fig:ConcG2}(a)  for $m=1,n=2$.
The behavior  for $\gamma=0$ is as shown previously, as $|\gamma|$ increases the entanglement increases as expected because the overlap between the two parties of the bipartite 
states diminishes. In  Fig. \ref{fig:ConcG2}b we consider the case with $m=0, n=20$. For $\gamma=0$ a wider region of maximal entanglement seen from $C(\psi^{AB})$ can be obtained compared to the case $m=1,n=2$. Qualitatively (a) and (b) have a similar behavior, which is jus a $\gamma$ displacement effect of the region of null entanglement.
This also shows that  with $|m-n|\gg 1$ or $|\gamma|>0$, the superposition values for $u$ (and hence $v$) are less restrictive when looking for high values of entanglement. 
These results also show similarities with Fig. \ref{fig:comparison}(b) and \ref{fig:comparison}(e) for $u=1/\sqrt{2}$ as the  states involved have non vanishing
 overlap for all $\gamma$ values, i.e. $\bra{10+\gamma}\hat{a}^{m}\hat{a}^{\dagger n} \ket{10-\gamma}\ne 0$.

%
%

\subsection{Depolarizing channel effects}
%
In this section we include the effect of decoherence of the state in Eq. (\ref{eq:estadoAB}) by means of a simple depolarizing channel \cite{Nielsen00a,Holevo12}. This mechanism maps the density matrix $\rho_0=\ket{\psi^{AB}}\bra{\psi^{AB}}$ into a mixture of the original $\rho_0$ and the completely mixed state $I/d$, with $d=\text{dim} (\rho)=4$ in the form $\rho = (1-\lambda)\ket{\psi^{AB}}\bra{\psi^{AB}} + \frac{\lambda}{d}I $. This map has a consistent Kraus operators representation, which assures a completely positive and trace-preserving map. For convenience, we have rewritten the density matrix with a $p$ parameter in the form $\lambda/d=p/3 \Rightarrow \lambda = 4p/3$. This is equivalent to study the entanglement of a mixed state density matrix with the form
\begin{equation}
\rho=\frac{p}{3}\text{I}+\left( 1-\frac{4p}{3} \right)\ket{\psi^{AB}}\bra{\psi^{AB}},
\label{eq:Ens}
\end{equation}
where $\ket{\psi^{AB}}$ is the entangled state Eq. (\ref{eq:EstadoAB_BaseON}) (or equivalently Eq. (\ref{eq:estadoAB})), and $ p $ is the mixing parameter. It is important to note that the full depolarizing effect occurs for $p=3/4$ with the full mixture of the density matrix and that the pure state is recovered for $p=0$. One may also note that if the state $\ket{\psi^{AB}}$ is maximally entangled, Eq. (\ref{eq:Ens}) is equivalent to a Werner state for which the upper bound of the mixing parameter is $p=1/2$ before the entanglement vanishes, and also Eq. (\ref{eq:Ens}) has the effect of an entanglement breaking channel \cite{Audretsch}. This case, serves as a reference for the extremal values of $p$ before the entanglement becomes null for maximally, non-maximally and non entangled states that can be obtained from Eq. (\ref{eq:EstadoAB_BaseON}) upon different combinations of parameters $\{u,v,\alpha,\beta,m,n\}$. Then one would expect that the extremal value of $p$ obtained for the maximally entangled state is an upper bound for all the other non-maximally entangled states. The effect of adding photons on the extremal values of $p$ before the entanglement vanishes can also be studied and is discussed in the next subsections. The concurrence of Eq. (\ref{eq:Ens}) is calculated with the Wootters formula as  \cite{Wootters1998},
\begin{equation}
 C(\rho)= \max \{0,\lambda_1-\lambda_2-\lambda_3-\lambda_4\},
 \label{eq:ConcW}
\end{equation}
where $\lambda_i$ are the square roots of the eigenvalues of the auxiliary matrix 
$\rho \tilde{\rho}$, where $\tilde{\rho}=(\sigma_y \otimes \sigma_y) \rho^* (\sigma_y \otimes \sigma_y) $ is the density matrix after application of the time reversal operator for two qubits, $\rho^*$ is the complex conjugate of the density matrix and
$\sigma_y$ is the y-Pauli matrix. To be clear in the definition of $\rho^*$, let $\rho$ be expanded in the computational basis as $\rho=\sum_{ \{i,j,k,l\}=0 }^{1} \rho_{ijkl} \ket{i,j}\bra{k,l}$, hence the complex conjugation of $\rho$ is understood as $\rho^*=\sum_{\{i,j,k,l\}=0}^{1} \rho_{ijkl}^* \ket{i,j}\bra{k,l}$.

\begin{figure}
	\centering
	\includegraphics[width=0.75\textwidth]{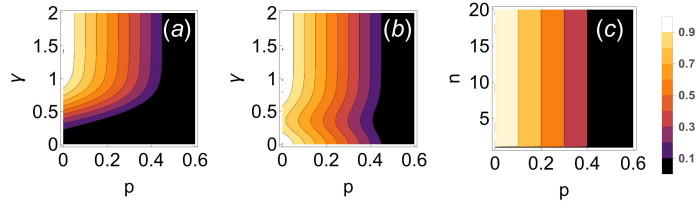}
	\caption{Concurrence for the ensemeble in Eq. (\ref{eq:Ens}) vs $p$ for the state of Eq. (\ref{eq:edocgamma}) with and the conditions: (a) $\alpha=\beta=0,m=n=0,u=v=1/\sqrt{2}$ vs $\gamma$, (b) $\alpha=\beta=0,m=1,n=2,u=v=1/\sqrt{2}$ vs $\gamma$ and (c)	$\alpha=\beta=3,\gamma=0,m=1,u=-v=1/\sqrt{2}$ vs the photon added number $n$.}
	\label{fig:ConcWN}
\end{figure}

We are interested in this section in determining the behavior of $p_{\text{crit}}$ of null entanglement for the different parameters of our entangled PACS, 
namely $( \alpha,\beta, \gamma, m,n)$. In Fig. \ref{fig:ConcWN} we examine the behavior $C(\rho)$ for different values of $p$ for the state in Eq. (\ref{eq:edocgamma}) for ECS, 
i.e., where the phases of of CS are opposite or $u=-v=1/\sqrt{2}$ as a comparison. The concurrence is determined for the following conditions: Fig. \ref{fig:ConcWN}(a) Concurrence as a function $p$ and $\gamma$  for $\alpha=\beta=0$, $u=v=1/\sqrt{2}$  with no added photon $m=n=0$. The critical values of $p$ 
for null entanglement  are $\gamma$ dependent; for larger $\gamma$ the larger values of  $p$, until it saturates to a value $p\approx 1/2$. In Fig. \ref{fig:ConcWN}(b)
  we show the case $\alpha=\beta=0$, $u=v=1/\sqrt{2}$  with $m=1,n=2$ of one photon added. These states with opposite phases exhibit an almost constant behavior of concurrence 
  for all values of $\gamma$, except for $\gamma \approx 0.36$ which is related to the maximum overlap $\bra{\gamma} \hat{a}^{ 2} \hat{a}^\dagger\ket{-\gamma}$. 
  For $\gamma=0$ or larger $\gamma$ the overlap tend to vanish which makes the states more distinguishable  and therefore produces a larger concurrence.
   In Fig. \ref{fig:ConcWN}(c) $C(\rho)$ is presented for $\alpha=\beta=3$ considering $\gamma=0$ with $u=-v=1/\sqrt{2}$, $m=1$ and varying $n$ and $p$. This figure shows that concurrence for the case $u=-v=1/\sqrt{2}$ do not depend on the difference on added photons $|m-n|$ but only on the value of $p$.

\begin{figure}
\centering
\includegraphics[width=0.75\textwidth]{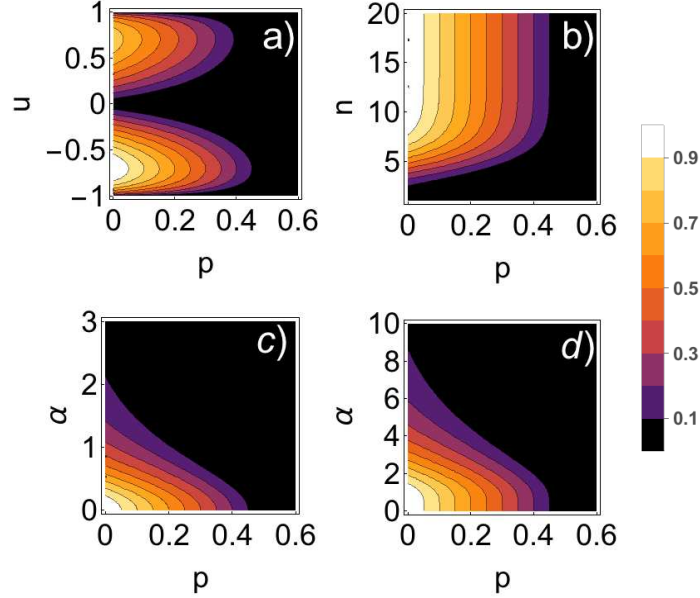}
\caption{Effects of the depolarizing channel to the variation of $p$ for (a) $\alpha=\beta=3,m=0,n=5$ vs. $u$, (b) $\alpha=\beta=3,m=0,u=v=1/\sqrt{2}$ vs. $n$, (c) $u=v=1/\sqrt{2},m=0,n=1,\alpha=\beta$ vs. $\alpha$ , (d)  $u=v=1/\sqrt{2},m=1,n=5,\alpha=\beta$ vs. $\alpha$.}
\label{fig:ConcWP}
\end{figure}

 In Fig. \ref{fig:ConcWP} we present contour curves $C(\rho)$ for different scenarios where the concurrence is highly dependent on all parameters of the state 
 in Eq. (\ref{eq:estadoAB}) and $p$ from Eq. (\ref{eq:Ens}). In Fig. \ref{fig:ConcWP}(a) concurrence as a function of $p$ and $u$ (and hence $v$) $C(p,u)$ is shown for a state with $\alpha=\beta=3, m=0,n=5$ fixed. The expected behavior for $v=-u=1/\sqrt{2}$ with the largest concurrence is obtained, and also large 
 values are obtained in the vicinity of $u=v=1/\sqrt{2}$, as well. In  Fig. \ref{fig:ConcWP}(b) we present the effect of added photon on concurrence $C(n,p)$, 
 here $\alpha=\beta=3,m=0$ and $u=v=1/\sqrt{2}$ are fixed, while the number $n$ of added photons in $B$ mode are varied as well as $p$ from the ensemble in
  Eq. (\ref{eq:Ens}). The figure shows how the concurrence contours are more robust to lower values of $p$ and larger values of $n$. This occurs because the 
  term $zN_2$ from Eq. (\ref{eq:EstadoAB_BaseON}) decreases as the difference $|m-n|$ increases. This can be seen as a reduction in the effective overlap 
  of the states $\hat{a}^{\dagger m}\ket{\alpha}$ and $\hat{a}^{\dagger n}\ket{\alpha}$, which make them more distinguishable. In figures \ref{fig:ConcWP}(c)-\ref{fig:ConcWP}(d)
   the concurrence $C(\alpha,p)$ is plotted for  positive superpositions $u=v=1/\sqrt{2}$ for fixed $m,n$ varying $\alpha=\beta$ and $p$. Specifically in Fig. \ref{fig:ConcWP}(c)
    is for  $m=0,n=1$ and in  Fig. \ref{fig:ConcWP}(d) for  $m=1,n=5$. These conditions  show the dependence on the concurrence for a given $\alpha$ and the difference on the added photons $|m-n|$ vs $p$. The behavior for both panels is qualitatively the same, and only corresponds to a different scale in $\alpha=\beta$, i.e., $C(\rho)$ decreases with $\alpha$ growing, and the values of $C(\rho)$ depend on the difference $|m-n|$. For all condition $C(\rho)$ vanishes for $p>0.5$.

\subsection{Critical values of p}
\begin{figure}
	\centering
	\includegraphics[width=0.8\textwidth]{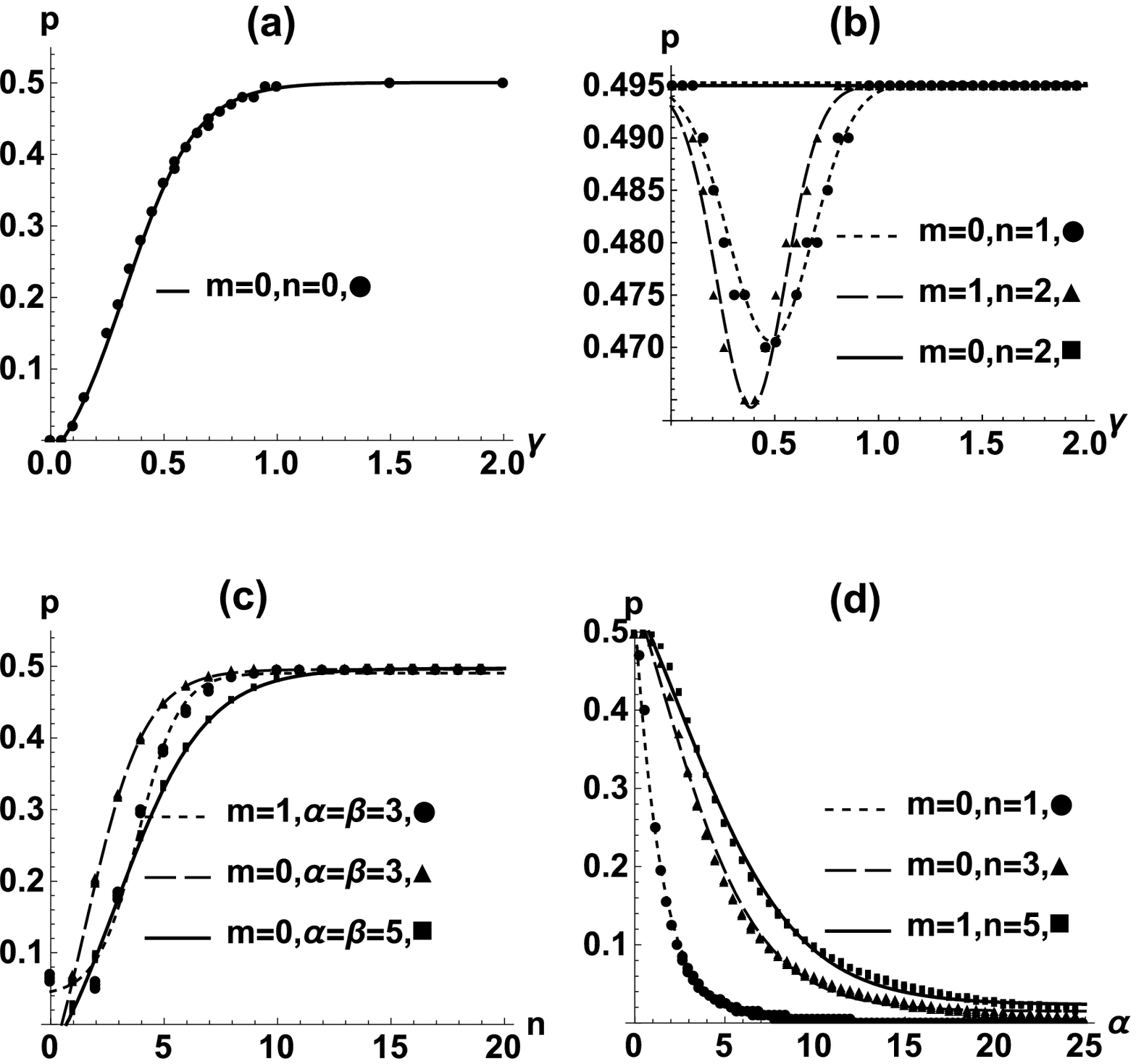}
	\caption{Critical values $p_{\text{crit}}$ for which the ensemble Eq. (\ref{eq:Ens}) becomes unentangled for different states $\ket{\psi}$. (a) $\ket{\psi} = N( \ket{\gamma}\ket{-\gamma}+\ket{-\gamma}\ket{\gamma} )$, (b)  $\ket{\psi} = N ( \hat{a}^{\dagger m}\otimes \hat{b}^{\dagger n}\ket{\gamma}\ket{-\gamma}+\hat{a}^{\dagger n}\otimes \hat{b}^{\dagger m}\ket{-\gamma}\ket{\gamma} )$, (c)-(d)  $\ket{\psi} = N ( \hat{a}^{\dagger m}\otimes \hat{b}^{\dagger n}\ket{\alpha}\ket{\alpha}+\hat{a}^{\dagger n}\otimes \hat{b}^{\dagger m}\ket{\alpha}\ket{\alpha} )$ with $m, \alpha$ fixed in (c) and $m,n$ fixed in (d). For all, $N$ is a suitable normalization constant.}
	\label{fig:ajustes}
\end{figure}

To find the critical values of $p$ for which the entanglement vanishes $p_{\text{crit}}$, we analyze the behavior of the concurrence as a function of the parameter $p$ for
 different superpositions of PACS. It may be noted that $p_{\text{crit}}$ works as an indicator of the robustness of  the entanglement for the state $\ket{\psi}$, where for each case considered we have dropped the superscript $AB$ for simplicity. The larger $p_{\text{crit}}$ is, the more difficult is to destroy the entanglement and therefore the need of a higher weight to produce the full mixture $I$ of Eq. (\ref{eq:Ens}).
The critical points are numerically fitted to a simple curve to have a better estimation of the bound of $p$ that keeps the entanglement properties of the mixed state.
As a reference, the state in Eq. (\ref{eq:ECS}) is analyzed. In Fig. \ref{fig:ajustes}(a) we show the critical $p$ values as a function of $\gamma$, clearly it increase from 0 to 1/2 as the average photon number is increased. This curve can be fitted by a hyperbolic function (continuous curve) of the form
\begin{equation}
	p_{\text{crit}} (\gamma)= a + b \tanh (d (\gamma-c)),
	\label{eq:tanhadj}
\end{equation}
With the fitting parameters $(a,b,d,c)$ presented in table \ref{tb:one} (first row).  This shows that these states become stronger when $\gamma$ increases. 
On the other hand, if $n,m$ photons are added to each corresponding mode, resulting in Eq. (\ref{eq:edocgamma}), things are quite different as it can be observed 
in Fig. \ref{fig:ajustes}(b) that for $\gamma \rightarrow 0$, $p_{\text{crit}} \rightarrow 0.495$ at almost constant value with a slight $6\%$ decrease for the cases when $|m-n|=1$
 at $\gamma \approx 0.5$, and the corresponding increment to the asymptotic value of $0.495$. A constant value of $p_{\text{crit}}=0.495$ is reached for the case $m=0,n=2$, 
 which persists for $|m-n|>1$.Effect that confirms that photon addition makes entangled states more robust, even for a small average photon number in the channel.

\begin{table}
	\caption{Hyperbolic fitting parameters of the critical $p_{\text{crit}}$ as a function of $variable=(\gamma \text{ or } n)$, for different entangled PACS}
\label{tb:one}
	\begin{tabular}{*6c}
		\toprule
		State & Fig  & a & b & c & d\\
		\midrule
		\multicolumn{6}{l}{ $\ket{\psi}=N (\ket{\gamma}\ket{-\gamma}+\ket{-\gamma}\ket{\gamma})$}, \\ \cline{2-6}
		{} & \multicolumn{5}{l}{$\alpha=0, \beta = 0, \gamma = variable, m=0, n=0$}\\ \cline{2-6}
		{} & \ref{fig:ajustes}(a)      & $0.2082 $    & $0.2925  $& $0.3275 $ & $3.2087 $ \\
		{} & {} & $\pm 5.742\e{-3}$ & $\pm 7.37\e{-3}$ & $\pm  7.792\e{-3}$ & $\pm  1.207\e{-1}$  \\ \cline{2-6}
		\multicolumn{6}{l}{ $\ket{\psi}=N (\hat{a}^{\dagger m}\otimes\hat{b}^{\dagger n}\ket{\alpha}\ket{\alpha}+\hat{a}^{\dagger n}\otimes\hat{b}^{\dagger m}\ket{\alpha}\ket{\alpha})$}\\ \cline{2-6}
		{} & \multicolumn{5}{l}{$\alpha=3, \beta = 3, \gamma = 0, m=0, n=variable$}\\ \cline{2-6}
		{} &  \ref{fig:ajustes}(c)    & $0.1355 $    & $0.3602  $& $1.5161 $ & $0.383228 $ \\
		{} & {} & $\pm 7.83\e{-3}$ & $\pm 7.97\e{-3}$ & $\pm  6.01\e{-2}$ & $\pm  5.64\e{-3}$  \\ \cline{2-6}
		{} & \multicolumn{5}{l}{$\alpha=3, \beta = 3, \gamma = 0, m=1, n=variable$}\\ \cline{2-6}
		{} &  \ref{fig:ajustes}(c)    & $0.1666 $    & $0.3306  $& $2.8647 $ & $0.2549 $ \\
		{} & {} & $\pm 4.21\e{-3}$ & $\pm 4.5\e{-3}$ & $\pm  5.78\e{-3}$ & $\pm  3.54\e{-3}$  \\ \cline{2-6}
		{} & \multicolumn{5}{l}{$\alpha=5, \beta = 5, \gamma = 0, m=0, n=variable$}\\ \cline{2-6}
		{} &\ref{fig:ajustes}(c)       & $0.2651 $    & $0.2255  $& $3.8557 $ & $0.5551 $ \\
		{} & {} & $\pm 5.48\e{-3}$ & $\pm 6.08\e{-3}$ & $\pm  8.44\e{-2}$ & $\pm  4.3\e{-2}$  \\ \cline{2-6}		
				{} & \multicolumn{5}{l}{$\alpha= \beta = variable, \gamma = 0, m=0, n=1$}\\ \cline{2-6}
				{} &  \ref{fig:ajustes}(d)      & $.6684 $    & $-0.6655  $& $-0.5456 $ & $0.4163 $ \\
				{} & {} & $\pm 0.0959$ & $\pm 0.0961$ & $\pm 0.2927$ & $\pm  0.0162$  \\ \cline{2-6}		
				{} & \multicolumn{5}{l}{$\alpha=\beta = variable, \gamma = 0, m=0, n=3$}\\ \cline{2-6}
				{} & \ref{fig:ajustes}(d)    & $0.4473 $    & $-0.4323  $& $1.2726 $ & $0.18 $ \\
				{} & {} & $\pm 2.6\e{-2}$ & $\pm 2.66\e{-2}$ & $\pm  3.34\e{-2}$ & $\pm  6.03\e{-3}$  \\ \cline{2-6}		
				{} & \multicolumn{5}{l}{$\alpha=\beta = variable, \gamma = 0, m=1, n=5$}\\ \cline{2-6}
				{} & \ref{fig:ajustes}(d)     & $0. 0.4025 $    & $-0.3793  $& $2.4903 $ & $0.1531 $ \\
				{} & {} & $\pm 1.69\e{-2}$ & $\pm 1.76\e{-2}$ & $\pm  3.05\e{-1}$ & $\pm  4.8\e{-3}$  \\ \cline{2-6}		
		\bottomrule
		\end{tabular}

\end{table}
The data of critical probability for $\alpha= \beta=0$ with photon added can be fitted by a Gaussian shaped function( negative amplitude $b<0$) as function of $\gamma$

	\begin{equation}
	p_{\text{crit}} (\gamma)=a + b \exp \left( \frac{-(\gamma-c)^2}{v^2}   \right)
	\label{eq:ajusteGauss}
	\end{equation}

where the values for the parameter $(a,b,d,c)$ are  presented in table (\ref{tb:dos}). The first row is
for  $m=0,n=1$ and the second row the case $m=1,n=2$. $p_{\text{crit}}$ for $m=0, n=2$ is also adjusted to a Gaussian function, but $b$ is too small to be 
observed resulting in the constant behavior shown. A difference of two or more
photons added makes the states, even for smaller values of $\gamma$, robust to the depolarizing effect.

If we consider  states with $\gamma =0$ but $\alpha=\beta>0$ the entanglement is lower for a given amount of added photons $m,n$; however, this mixed system can gain entanglement
if the difference $|m-n|$ is increased, as shown in previous section for the pure state. Now, the same reasoning applies for finding the $p_{\text{crit}}$ of these states.
The state under analysis is then
$\ket{\psi} = (N/\sqrt{2}) (\hat{a}^{\dagger m} \otimes \hat{b}^{\dagger n}\ket{\alpha}\ket{\alpha} + \hat{a}^{\dagger n} \otimes \hat{b}^{\dagger m}\ket{\alpha}\ket{\alpha} )$
 where $p_{\text{crit}}$ is shown in Fig. \ref{fig:ajustes}(c),
considering $m=0$ we see that for $\alpha$ small and n=0, the critical values of $p$ are zero with no surprise, because this state has null entanglement.
When $n \geq 1$, $p_{\text{crit}}>0$, the plots in the figure show that for a given combinations of $\alpha=\beta$ and $m$, $p_{\text{crit}}$ increases with $n$, which shows
again that robustness can be gained by the process of adding more photons to one mode while letting the other fixed. They all reach the asymptotic value of $p_{crit}=1/2$.  All the $p_{\text{crit}} (n)$ data are also fitted to
a hyperbolic tanh function in the form of $ p_{\text{crit}} (n)= a + b \tanh (d (n-c))$. The results are in table \ref{tb:one}, for the parameters $(a,b,d,c)$. In the second row are for $\{m=0,\alpha=3\}$
the third row for  $\{m=1,\alpha=3\}$ and the fourth row for $\{m=0,\alpha=5\}$. We will show in the next section that the state with $p_{crit}=1/2$ correspond a states with maximum entanglement.

Finally, in Fig. \ref{fig:ajustes}(d) the $p_{\text{crit}}(\alpha)$ as a function of the coherent state $\alpha$, is shown for combinations of $m,n$ for different  $\alpha=\beta$.
It is clear that the critical values will decrease with $\alpha$ as the state of the ensemble is less entangled, and tend asymptotically to zero for $\alpha$ large.
These data are also well fitted by the hyperbolic tan function $ p_{\text{crit}} (\alpha)= a + b \tanh (d (\alpha-c))$, with amplitude $b<0$. Eq. (\ref{eq:tanhadj}). The fitting parameters are also presented in table \ref{tb:one}.
The fifth row is for $\{m=0,n=1\}$, the sixth row is for $\{m=0,n=3\}$ and seventh for $\{m=1,n=5\}$. These result are in agreement with the fact that for $|m-n| \gg 1$ there are more regions of the coherent state value $\alpha$ with noticeable entanglement.
\begin{table}
	\caption{Gaussian fitting parameters of the critical $p_{\text{crit}}$ as a function of $\gamma$, for different entangled PACS}
\label{tb:dos}
	\begin{tabular}{*6c}
	\toprule
	State & Fig  & a & b & c & v\\
	\midrule
	\multicolumn{6}{l}{ $\ket{\psi}=N (\hat{a}^{\dagger m}\otimes\hat{b}^{\dagger n}\ket{\gamma}\ket{-\gamma}+\hat{a}^{\dagger n}\otimes\hat{b}^{\dagger m}\ket{-\gamma}\ket{\gamma})$}\\ \cline{2-6}
	{} & \multicolumn{5}{l}{$\alpha=0, \beta = 0, \gamma = variable, m=0, n=1$}\\ \cline{2-6}
	{} & \ref{fig:ajustes}(b) & $0.4952 $ & $-0.02459  $& $0.479 $ & $0.2824 $ \\
	{} & {} & $\pm 3.5\e{-4}$ & $\pm  7.18\e{-4}$ & $\pm  6.78\e{-3}$ & $\pm  1.07\e{-2}$  \\ \cline{2-6}
	{} & \multicolumn{5}{l}{$\alpha=0, \beta = 0, \gamma = variable, m=1, n=2$}\\ \cline{2-6}
	{} &  \ref{fig:ajustes}(b) & $0.4952 $ & $-0.03074  $& $0.3859 $ & $0.2333 $ \\
	{} & {} & $\pm 3.43\e{-4}$ & $\pm  8.84\e{-4}$ & $\pm  4.98\e{-3}$ & $\pm  8.07\e{-3}$\\ \cline{2-6}
	{} & \multicolumn{5}{l}{$\alpha=0, \beta = 0, \gamma = variable, m=0, n=2$}\\ \cline{2-6}
	{} &  \ref{fig:ajustes}(b) & $0.4952 $ & $-8.07\e{-17}  $& $1.0045 $ & $1.019 $ \\
	\bottomrule
	\end{tabular}
\end{table}
The behavior for all the critical values of $p$ confirm that the process of adding photons acts as a robustness enhancer of the
entanglement properties of the state in both scenarios of pure a mixed states.

\subsection{The critical bound $p_{\text{crit}}=1/2$}
In order to provide insight into the critical asymptotic value $p_{\text{crit}}=1/2$, we consider the family of maximally entangled photon added coherent states (EPACS) entangled states that occurs for photon added coherent
states under consideration.The known ECS state $\ket{\psi}=N(\ket{\gamma}\ket{-\gamma}-\ket{-\gamma}\ket{\gamma})$ has concurrence $C=1$. This permits to obtain the limiting
values for $p$ that destroy entanglement, as no further photon addition operations could increase this value of entanglement. By starting with this state, the
density matrix $\rho_0=\ket{\psi}\bra{\psi}$ in the computational basis is

	\begin{equation}
	\rho_0=	\left(
		\begin{array}{cccc}
		0 & 0 & 0 & 0 \\
		0 & \frac{1}{2} & -\frac{1}{2} & 0 \\
		0 & -\frac{1}{2} & \frac{1}{2} & 0 \\
		0 & 0 & 0 & 0 \\
		\end{array}.
		\right)
		\label{eq:GMalphaminusalpha}
	\end{equation}
It is straightforward to prove that $\rho_0$ is independent of the values $m,n$. This state is part of a larger family of maximally entangled photon added coherent states (MEPACS) under study, as the superposition with
$u=-v=1/\sqrt{2}$ for all $\alpha$ and $\beta$ when $m \ne n$, and for $\alpha \ne \beta$ if $n=m$ construct the MEPACS as
	\begin{equation}
		\ket{\psi} = N(1/2,1/2)(1/\sqrt{2})( \hat{a}^{\dagger m}\otimes \hat{b}^{\dagger n} \ket{\alpha}\ket{\beta} - \hat{a}^{\dagger n}\otimes \hat{b}^{\dagger m} \ket{\alpha}\ket{\beta}),
		\label{eq:edomax}
	\end{equation}
 creates the same density matrix in computational basis as Eq. (\ref{eq:GMalphaminusalpha}) as for the  case of  ECS $(\alpha=-\beta=\gamma, n=0,m=0)$,  and with same entanglement properties. With a simple substitution,  the constant $N$ is
\begin{equation}
	\begin{aligned}
		N(1/2,/12)&=\Big[ L_m(-|\alpha|^2)L_n(-|\beta|^2)m!n!-  e^{-(|\alpha|^2+|\beta|^2)}|\beta|^{2(m-n)}  \\
		&\quad \times (\Gamma(1+m))^2 |{}_1\tilde{F}_1 (1+m;1+m-n;\alpha^*\beta  )|^2   \Big]^{-1/2},
	\end{aligned}
	\label{eq:NTo1}
\end{equation}
In the computational basis, this states produces a maximal entangled Bell state according to Eq. (\ref{eq:Concurrence_Estado_AB}), $C=1$ due to the fact that in this
case  $N(1/2,1/2)=N_{1}N_{2}$
Therefore Eq. (\ref{eq:GMalphaminusalpha}) is used to construct the mixed state Eq. (\ref{eq:Ens}) for these family of states , resulting in the matrix
	\begin{equation}
	\rho=\frac{p}{3}I + \left(1-\frac{4p}{3}\right)\rho_0	=\left(
		\begin{array}{cccc}
		\frac{p}{3} & 0 & 0 & 0 \\
		0 & \frac{1}{6} (3-2 p) & \frac{1}{6} (4 p-3) & 0 \\
		0 & \frac{1}{6} (4 p-3) & \frac{1}{6} (3-2 p) & 0 \\
		0 & 0 & 0 & \frac{p}{3} \\
		\end{array}
		\right),
	\end{equation}
This has a $X$ form
	\begin{equation}
		\rho=\left(
		\begin{array}{cccc}
		u_{+} & 0 & 0 & 0 \\
		0 & w & z & 0 \\
		0 & z & w & 0 \\
		0 & 0 & 0 & u_{-} \\
		\end{array}
		\right),
	\end{equation}
 where the concurrence $C$ can be calculated and is expressed  in the compact form \cite{connor2001},

\begin{equation}
C=2\max\biggr(0, \vert z \vert-\sqrt{u^{+}u^{-}}\biggr).
\label{equation,21}
\end{equation}
For our case, we have $z= 2p/3-1/2$ and $u_{+}=u_{-1}= p/3$.  The concurrence is then given by
	\begin{equation}
		C( p )  = \text{max}\{0, 1-2p\},
		\label{eq:limCp}
	\end{equation}
which sets  value $p=1/2$ as the extreme value of $p$, denoted as $p_{\text{crit}}$ before concurrence becomes zero in the
ideal situation  of a state with maximum entanglement in a depolarizing channel. It is important to note that any linear
combination state different to Eq. (\ref{eq:edomax}) will result in lower values for $p_{\text{crit}}$ as the state will have lower entanglement.
This is the reason for which further calculations are carried out on the positive linear combination, i.e., where $u=v=1/\sqrt{2}$, as any variation on
the parameters of the state Eq. (\ref{eq:edocgamma}) changes $p_{\text{crit}}$. More important are the cases where initial $p_{\text{crit}}$ is low
and after photon addition operations this value can be increased.
This increment, can be easily understood if one starts with the trivial state
$\ket{\psi} \propto \ket{\alpha}\ket{\alpha}+\ket{\alpha}\ket{\alpha} \propto \ket{\alpha}\ket{\alpha}$,
that obviously has null entanglement, and compares it with the result of mixing $\hat{a}^{\dagger}\ket{\sqrt{2}\alpha}$
with $\ket{0}$ in a $50/50$ beamsplitter obtaining a state of the form $\ket{\psi} \propto \hat{a}^\dagger\ket{\alpha}\ket{\alpha}+\ket{\alpha}\hat{b} ^\dagger\ket{\alpha}$
whose entanglement now is not null. The only difference in the generation of one state and the other is the photon addition
process that is experimentally feasible \cite{Zavatta2004,Jeong2014}.
The increments of $p_{\text{crit}}$ in the end are related to the initial
 degree of entanglement of $\rho_0$. Therefore, if photon addition increases
 $C(\rho_0)$ the needed value of $p$ to destroy the entanglement will increase as well.

\section{Conclusions}
We have presented a study about the entanglement properties of an arbitrary superposition of $n,m$
photon added coherent states of the form of Eq. (\ref{eq:estadoAB}) and the conditions to obtain maximal entanglement were
presented. We evaluate the influence of the difference $|n-m|$ and the values of the product $\alpha \beta$ in the concurrence amount.
It was found that in addition to the standard form of ECS where coherent states with opposite phases are used to obtain larger values of entanglement,
a large difference $|m-n| \gg 1$ could be used with a similar effect as well, which implies that photon addition on positives superpositions of CS enhance the entanglement properties of the system. The largest values of entanglement are obtained around two linear combinations that occur
for $u=\pm 1/\sqrt{2}$, and the exact value of $u$ could be unstrained if the coherent states have opposite phases or if the amount of added photons in each mode of interest is very different. We have also discussed the effects of a depolarizing channel to the entanglement properties of the superposition and have found that the boundary of $p$ for non vanishing $C(\rho)$ is $p \to 0.5$ from the left, and we provide analytical results to support that numerical finding, regardless of the configuration of the superposition. However, $C(\rho)$ can vanish for lower values of $p$ depending on the characteristics of the state.
Numerical fitting of the critical probability $p_{crit}$ are given an we show a  hyperbolic relation with $\gamma$ and added photon  number $n$ .
We also obtained a family of PACS superpositions with maximal entanglement Eq.(\ref{eq:edomax}), without the need of a negative phase in the  coherent state. 
These states have  the maximum critical probability $p_{crit}=1/2$ under depolarization effects and could be useful in quantum information processing tasks with continuous variables.

\section{Acknowledgments}
FDS acknowledges receipt of a PhD scholarship from CONACYT (Grant No. 331668). Thanks Dr. E. Cota
for useful comments on the manuscript.


\end{document}